%% file: RASproc00hepex.tex
\documentclass[a4paper,12pt,oneside]{article}
\usepackage[english]{babel}
\usepackage{rotating}
\usepackage{epsfig}
\usepackage[centertags]{amsmath}
\usepackage{amssymb,amscd,./styles/amsatdef}
\usepackage{array}
\usepackage{multirow}
\usepackage{supertabular}
\usepackage{colordvi}
\usepackage{./styles/axodraw}
\usepackage{dcolumn}
\usepackage{./styles/mcite}
\usepackage{cite}
\usepackage{xspace}
\input ./com_def_tex/zeuspap_def.tex

\input ./com_def_tex/zeuspap_fmt.tex

\def\drftdate{{\today}}

\newcommand {\pom} {I\!\!P}
\newcommand {\pomsub} {{\scriptscriptstyle \pom}}

\input ./com_def_tex/charm_com.tex
\begin{document}
\selectlanguage{english}
\thispagestyle{empty}
\vskip2.mm
\vskip12.mm
\vskip2.2cm
\centerline{\bf\LARGE Measurement of $D^{*\pm}$ diffractive cross sections} 
\vskip.6cm
\centerline{\bf\LARGE in photoproduction at HERA}
\vskip1.0cm
                     
\centerline{\large {\bf I.A.Korzhavina}\footnote[1*]{Institute of Nuclear Physics, 
Moscow State University, Russia \\ \hspace*{-.3cm}$^*$ email: irina@mail.desy.de}
} 
\centerline{(for  the ZEUS Collaboration)}  
\vskip0.2cm
\date{\drftdate}
%

%
%
 
\renewcommand{\baselinestretch}{1.0}
\large

\vskip1.5cm
\centerline{
  \begin{minipage}{15.cm}
    \hspace*{\parindent}
\phantom{.....} The first measurement of $D^{*\pm}$ meson diffractive 
photoproduction cross sections has been performed  
with the ZEUS detector at the HERA $ep$ collider, 
 using an integrated luminosity of 38$\pbi$. The measurement has been
   performed for photon--proton center-of-mass energies in the range
 \mbox{$130 < W < 280$ GeV} ~ and photon virtualities $Q^2 < 1$ \GeVs. 
 $D^{*\pm}$ mesons have been reconstructed with $p_T^{D^*}>2$ GeV 
 and \mbox{$-1.5<\eta^{D^*}<1.5$} from the decay channel 
 $D^{*+} \to D^0 \pi^+_s$ with $D^0 \to K^-\pi^+$ (+c.c.). The diffractive 
 component  has been selected with \mbox{$0.001<x_{\pomsub}<0.018$}. 
  The measured cross section in this kinematic range is: 
 \mbox{$\sigma_{ep \to e^{\prime}D^*Xp^{\prime}}^{diff} = 
 0.74 \pm 0.21 (stat.)^{+0.27}_{-0.18} (syst.) \pm 0.16 (p. diss.)$ ~nb}
 (ZEUS preliminary). 
Measured integrated and differential cross sections have been compared to
theoretical expectations.
  \end{minipage}
}

\vfill

\thispagestyle{empty}
%
 

%
%
\section{Introduction}
%
%

\hspace*{\parindent}
\phantom{.....} Charm production processes are the ones
 which proceed mainly through gluon--initiated hard subprocesses and are 
 perturbatively calculable. Thus, the diffractive production of charmed 
 mesons can  provide new tests of the partonic structure of diffractive 
 interactions, in  particular of their gluon component.

\parindent1.cm
During the years of the HERA collider operation, integrated and differential
cross sections for inclusive charm production were measured in
kinematic ranges where an effective signal separation from suppressed 
backgrounds could be achieved \cite{ZEUS-D*-incl,H1-D*-incl}. 
The measured cross sections for photoproduction (PhP) and deep
inelastic scattering (DIS) processes were compared with different
next-to-leading (NLO) pQCD calculations. DIS data were found to be in good
agreement with the calculations. The calculated PhP cross sections are lower
than the measured ones, especially in the forward (proton) direction. 

\parindent1.cm
As for the diffractive charm production, there are only preliminary results on
diffractive dissociation in DIS,
 measured with $D^{*\pm}$ mesons \cite{ZEUS-D*-disdif, H1-D*-disdif}.
Here we present preliminary results on measurements by the ZEUS Collaboration 
of  cross sections for 
diffractive photoproduction of \mbox{ $D^{*\pm}$(2010)}
 mesons\footnote{In the following, \mbox{ $D^{*\pm}$}(2010) will be referred 
                                                      to simply as $\Ds$.}
in the range of Pomeron fractional momentum \mbox{$0.001<x_{\pomsub}<0.018$}
at energies \mbox{$ 130 < W < 280 $} \Gev ~in the photon--proton center-of-mass 
frame ~ and  photon virtualities \mbox{$Q^2< 1$} \GeVs.
\mbox{$D^*$} mesons were reconstructed through the  decay channel
${\rm D}^{*+}\rightarrow {\rm D}^{0}\pi_s^{+} 
 \rightarrow ({\rm K}^{-}\pi^{+})\pi_s^{+}$  
(and c.c.) in the restricted kinematic region: 
\mbox{$p_T^{D^*} > 2$} \Gev ~and 
\mbox{$\vert \eta ^{ D^*}\vert  < 1.5 $}. Here $p_T^{D^*}$ is the $D^*$ meson
transverse momentum and $\eta ^{ D^*}=-$ln$($tan$(\theta/2)) $ is its pseudorapidity,
defined in terms of the $D^*$ polar angle $\theta $ with respect to the proton beam 
direction.

\parindent1.cm
The measurements were performed at the HERA collider with the ZEUS detector,
a detailed description of which can be found elsewhere \cite{ZEUS}.
The data were taken during 1996 and 1997, when HERA collided positron 
and proton beams with energies of 27.5 \Gev and 820 \Gev, respectively.
 An integrated luminosity of 38$\pbi$ was used for this measurement. 
Charged particles were measured in the central tracking detector (CTD) \cite{CTD}.
 To detect the scattered electron and to measure global energy values 
 the uranium-scintillator sampling calorimeter (CAL) \cite{CAL} was used. 
The luminosity was determined from the rate of the bremsstrahlung process 
$e^+p \to e^+\gamma p$, where the photon was measured by a lead scintillator
calorimeter \cite{LUMI}. 

%
%
\section{Kinematics of diffractive photoproduction}
%
%

\hspace*{\parindent}
We consider diffractive photoproduction in $ep$ scattering at HERA: 
\[ 
e(e) + p(p) \to e^{\prime }(e^{\prime }) + X + p^{\prime }(p^{\prime }),
 \]
where the four-momenta of particles are shown in brackets. The
collision occurs at the squared positron--proton center-of-mass energy 
$s=(e+p)^2$, and photon virtuality \mbox{$Q^2=-q^2$},   where
\mbox{$q=e-e^{\prime}$}.
The squared photon--proton center-of-mass energy 
$W^2=(p+q)^2$ is defined for this reaction.
One may consider that the interaction proceeds through a  
photon--Pomeron ($\pom$) scattering:
\[ \gamma(q) + \pom (P_{\pomsub}) \to  X, \]
 where 
 $P_{\pomsub}=p-p^{\prime}$. 
 This process  is described by
the invariant mass $M_X$ of the hadronic system X, produced by  
photon dissociation, and the fraction of the proton momentum
 \[ x_{\pomsub} = \frac{P_{\pomsub}\cdot q}{p\cdot q}\simeq \frac{M_X^2}{W^2},\]
 carried away by the Pomeron.

\parindent1.cm
The variables $W, M_X$ and $ x_{\pomsub}$ were reconstructed from the final 
hadronic system, measured by energy flow objects (EFO) \cite{EFO}, 
 made from tracks detected by the CTD and from energy deposits in the CAL cells.
The Jacquet--Blondel formula 
$W_{\JB} = \sqrt{2E_p\sum_i (E-P_z)_i}$ \cite{JB} 
was used to reconstruct $W$. Here $E_p$ is the proton beam energy. 
 The invariant mass of the diffractively produced system $M_X$ was calculated 
 with the formula \\
\mbox{$M_X^2 = \left(\sum_i  E_i   \right)^2 - \left(\sum_i P_{x_i}\right)^2
      - \left(\sum_i P_{y_i}\right)^2 - \left(\sum_i P_{z_i}\right)^2$}.
Sums in both equations run over energies $E_i$ and momenta $P_i$ of all EFOs. 
To select the W
range, $W_{\JB}$ was calculated with calorimeter  cell deposits  
only so as to be consistent with the inclusive charm photoproduction 
analysis \cite{ZEUS-D*-incl}.
 Measured values were corrected to the true ones 
by factors, determined from MC simulations of  diffraction as average ratios 
of reconstructed to generated values.  
All variables were reconstructed to an accuracy of better than 15\%.
 
%
%
\section{Event selection and $D^*$ reconstruction}
%
%
%

\hspace*{\parindent}
Event selection and $D^*$ reconstruction procedures are described 
in details elsewhere \cite{ZEUS-D*-incl}. Here a short description is given.
 
\parindent1.cm
Photoproduction events were selected by requiring that no scattered
positron was identified in the CAL~\cite{no-electron} and  the 
photon--proton center-of-mass energy
$W $ is between 130 and 280 \Gev.  Under these conditions, the
photon virtuality $ Q^{2} $ is limited to values less than  1 \GeVs.
  The corresponding median \( Q^{2} \) was estimated from a
Monte Carlo (MC) simulation to be about  $3 \times 10^{-4}$  \GeVs.
%
The $D^*$ mesons were reconstructed through the decay channel 
\mbox{$D^* \to (D^0\to K\pi)\pi_s$} by combining candidates from charged 
tracks measured by the CTD.
For the reconstruction, ``right charge'' track combinations, defined for
$(K\pi)$ with two tracks of  opposite charges
 and with a $\pi_s$ having the charge opposite to that of the $K$ meson in 
 the $(K\pi)$, were accepted as long as the combination of invariant masses  
 $\Delta M=M(K\pi\pi_s)-M(K\pi)$ 
 and $M(K\pi)$ are within wide mass-windows around the nominal
values of $\Delta M=M(D^*)-M(D^0)$ and $M(D^0)$ \cite{PDG}.
To determine the number of $D^*$ mesons in the signal, combinatorial 
background was modelled by ``wrong charge'' 
track combinations and subtracted after normalization to
the ``right charge'' distribution in the range $0.15 < \Delta M < 0.17$ \Gev. 
``Wrong charge'' combinations were defined for $(K\pi)$ with two tracks of the
same charge and with a $\pi_s$ of the opposite charge.
The measurements were performed in the pseudorapidity
range \( -1.5<\eta ^{D^*}<1.5 \), where the CTD acceptance is
high. 
  The kinematic region in $p_T^{D^*}$ was limited to \(2< p_T^{D^*}<8 \)~\Gev. 
 
\parindent1.cm
The MC event samples used for this analysis were prepared with 
the RAPGAP~\cite{RAPGAP},  
PYTHIA~\cite{PYTHIA} and HERWIG~\cite{HERWIG} generators. 
Diffractive interactions were modeled in the framework of the resolved Pomeron 
 model \cite{IPres} with $\beta (1-\beta)$ 
 or the H1 FIT2 \cite{IPfitH1} 
parametrisations for the initial partonic distributions in the Pomeron.
Here $\beta$ is the  fraction of the Pomeron momentum carried by a parton, that
couples to the Pomeron and participates in the hard interaction.
The MRSG~\cite{MRSG} and GRV-G~HO~\cite{GRV} parametrisations were used
for the proton and photon structure functions, respectively, when modelling
non-diffractive interactions. 
The fragmentation of the generated partons (parton shower evolution and 
hadronisation) was simulated according to the LUND model \cite{LUND} when
using  the RAPGAP or PYTHIA simulations. The HERWIG generator models the  
hadronisation process with a cluster hadronisation model.
The MC events were processed through the standard ZEUS detector and
trigger simulation programs and through the same event reconstruction
package as was used for data processing.  The shapes of MC and data
distributions were found to be in  reasonable agreement within statistical
errors. 

\parindent1.cm
Diffractive events were identified by a large rapidity gap (LRG) 
between the scattered proton, which
 escaped detection through the beam pipe, and the hadronic system X, 
produced by the dissociated photon. The LRG events were searched 
for using the $\eta_{max}$ method, for which $\eta_{max}$ was defined 
as the pseudorapidity of the most forward EFO
 with energy greater than 400 MeV. Fig.~1
 presents the $\eta_{max}$ distribution 
for all photoproduced $D^*$ mesons, reconstructed within the signal range \\
\mbox{$0.143 < M(K\pi \pi_s)-M(K\pi) < 0.148$ \Gev} ~and 
\mbox{$1.80 < M(K\pi) < 1.92$ \Gev} after the combinatorial background
subtraction.
This distribution shows two structures. The plateau-like structure at 
\mbox{$\eta_{max} \lesssim 2 $} is populated predominantly by the LRG events, 
while the wide
peak-like structure around $\eta_{max} \sim 3.5 $ originates from the
non-diffractive events and has an exponential fall-off towards lower values of 
$\eta_{max}$. From a comparison between the data points and a sum of simulated 
diffractive and non-diffractive event distributions, normalised to the data, 
  a cut-off of $\eta_{max}=1.75$ was chosen
 as a compromise between the magnitudes of the diffractive signal and
the non-diffractive background. 
The non-diffractive background fractions for subtraction were estimated from 
the MC-to-data 
distribution ratios, using non-diffractive MC simulations.

\parindent1.cm
When using the $\eta_{max}$ method for diffractive event selection, one needs to 
take into account the following properties of the method. The measurement of 
rapidities by the CAL is limited 
  to the edge of the forward beam hole of the CAL. Thus the proton 
 dissociative events, \mbox{$ep\to e^{\prime}XN$}, can satisfy the requirement 
 $\eta_{max}< 1.75$ if 
 the proton dissociative hadronic system $N$ has invariant mass small enough to
  pass undetected through the forward beam pipe.
 It was found earlier that the proton
 dissociation contribution comprises 0.31 $\pm$ 0.15 \cite{prodiss}. Measured
 cross sections were corrected for this value. A cut in $\eta_{max}$
 correlates with a range of accessible $x_{\pomsub}$ values.
$\eta_{max}< 1.75$ restricts \mbox{$x_{\pomsub} < 0.018$}. 
In addition, limited acceptance restricts \mbox{$x_{\pomsub}>0.001$}. 

%
%
\parindent1.cm
After the above selection and the ``wrong charge'' background subtraction 
a signal of \mbox{56 $\pm$ 10} diffractively photoproduced $D^*$ mesons was found
 in the $\Delta M$ distribution (Fig.2).

%
%
\section{Cross sections}
%
%

\hspace*{\parindent}
The inclusive $D^*$  production cross section is given by:
\[ \sigma_{ep\to D^* X}=\frac{N^{corr}_{D^*}}{\mathcal{L}\cdot 
    B_{D^*\to (D^0 \to K \pi)\pi }} ~ , 
\]
where $N^{corr}_{D^*}$ is the number of 
 observed $D^*$ mesons corrected for the acceptance, 
 \mbox{$ \mathcal{L}=38.0\pm 0.6 $} $\pbi$ is the integrated luminosity and 
\( B_{\scriptscriptstyle D^* 
\to (D^0 \to K \pi)\pi } =0.0263\pm 0.0010 \)
 ~is the combined
\mbox{\( D^* \to (D^0 \to K^+ \pi^-)\,\pi_s \)} decay branching ratio 
\cite{PDG}.
 Acceptance corrections were calculated using the RAPGAP MC sample.
 
\parindent1.cm
The total \(D^* \) diffractive photoproduction cross section 
in the kinematic region \mbox{$Q^2<1$ \GeVs}, \mbox{$130 < W < 280$ \Gev},
 \mbox{$ p_T^{D^*}>2$ \Gev}, \mbox{$ \vert \eta^{D^*}\vert < 1.5$} 
 and \mbox{$0.001 < x_{\pomsub} < 0.018$}  was measured  to be \\
\mbox{$ \sigma_{ep\to  e^{\prime}D^* Xp^{\prime}}^{diff}
 = 0.74 \pm 0.21 (stat.)^{+0.27}_{-0.18} (syst.) \pm 0.16 (p. diss.)$} nb
 (ZEUS preliminary). 
The last error is due to the  uncertainty  in the proton dissociative background
subtraction.     
Other sources of systematic uncertainties due to analysis and detector 
 features were studied and their effect on 
the cross section was estimated. The largest contributions to the systematic 
error came from the CAL energy scale uncertainty ($^{+12.0}_{-4.8}$\%), 
 the signal determination procedure  ($^{+16.4}_{-14.5}$\%), the  selection
 of diffractive events ($^{+11.3}_{-8.0}$\%) and  the acceptance correction
calculations ($^{+26.5}_{-16.9}$\%).   
 The overall normalisation uncertainties due to the error in the 
 luminosity value ($\pm 1.7\%$) and  in the $D^*$ ~and $D^0$ decay  branchings
 ($\pm 3.8\%$) were  not included in the systematic error quoted above.
All of the systematic uncertainties were added in quadrature to determine the
overall systematic uncertainty of $^{+35.6}_{-24.1}$\%.  
The summation of the systematic uncertainties was also performed for each 
bin of the differential distributions.
  
\parindent1.cm
The measured  $D^*$ diffractive photoproduction cross section, while only a
fraction of the total diffractive contribution, amounts 
$\sim 4$\% of the inclusive $D^*$ photoproduction cross section,  
  \( \sigma_{ep\to D^* X}= 18.9\pm  1.2(stat) ^{+1.8}_{-0.8}(syst) \) ~nb
  \cite{ZEUS-D*-incl},
measured in the same kinematic range. This fraction indicates that
 diffractive charm production is not suppressed as much as some early models
 predicted \cite{c-suppr}.
 
\parindent1.cm
 Measurements were compared to  resolved Pomeron model expectations
  \cite{IPres}, calculated with the RAPGAP Monte Carlo program
 in the same kinematic region.  Partonic distributions in the Pomeron were 
 parametrised by the fit to the HERA data
\cite{IPfitH1}, performed by H1 Collaboration (H1 FIT2). 
Only the BGF mechanism of charm production was accounted for.
      The leading order RAPGAP Monte Carlo, with the H1 FIT2 Pomeron
 parametrisation, predicts 1.42 nb for the $D^*$ diffractive photoproduction 
 cross section in the same kinematic range \cite{HJung}.
 
\parindent1.cm
 Differential cross sections for
$ d\sigma /dp_T^{D^*}$, $ d\sigma /d\eta ^{D^*} $, $ d\sigma /dM_X $ and
$ d\sigma /dx_{\pomsub} $ are presented in Figs.~3--6.
 All of the above mentioned systematic uncertainies were added in quadratures 
 with statistical errors (inner error  bars) in each bin to calculate the total 
 error (outer error  bars), both of which are  shown  in Figs.~3--6.

%
\parindent1.cm
 The measured differential cross sections  (Figs.~3--6), 
when  compared to the ones of the resolved Pomeron model
 calculated with the RAPGAP MC program
 show  reasonable agreement in shape with the theoretical expectations 
considering the measurement errors.
$d\sigma/dp_T^{D^*}$ agrees well and the other three distributions are shifted 
somewhat to larger values with respect to the predictions.

%
%
\section{Summary and conclusions}

\hspace*{\parindent}
The first measurement of diffractive $D^*$ photoproduction has been
performed with the ZEUS detector at HERA with a luminosity of
38$\pbi$. The preliminary results are reported
 here. The total \(D^*\) diffractive photoproduction cross section 
in the kinematic region \mbox{$Q^2<1$ \GeVs}, \mbox{$130<W<280$} $\Gev$,
 \mbox{$p_T^{D^*}>2$ \Gev}, \mbox{$\vert\eta^{D^*}\vert< 1.5$} 
 and \mbox{$0.001< x_{\pomsub}< 0.018$}  
 is measured  to be 
$ \sigma_{ep\to e^{\prime}D^* Xp^{\prime}}^{diff}=$ 
 \mbox{$0.74 \pm 0.21 (stat.)^{+0.27}_{-0.18} (syst.) \pm 0.16 (p. diss.)$} nb 
(ZEUS preliminary)
. 
 The leading order  calculations in the framework of the resolved Pomeron 
model predict 1.42 nb for this  cross section. 
The  differential cross section shapes for 
\mbox{$ d\sigma /dp_T^{D^*}$}, \mbox{$ d\sigma /d\eta ^{D^*} $}, 
\mbox{$ d\sigma /dM_X $} and \mbox{$ d\sigma /dx_{\pomsub} $} show 
 reasonable agreement  
 with the resolved Pomeron model considering the measurement errors.
%
%
\section{Acknowledgements}

\hspace*{\parindent}
We would like to thank the DESY Directorate for their strong support
and encouragement. The remarkable achievements of the HERA machine
group were essential for this work and
are greatly appreciated. We would like to thank H.Jung for fruitful
discussions and help in the use of the RAPGAP MC generator.

\newpage
\begin{figure}[tb]
\centerline{\hbox{
\psfig{figure=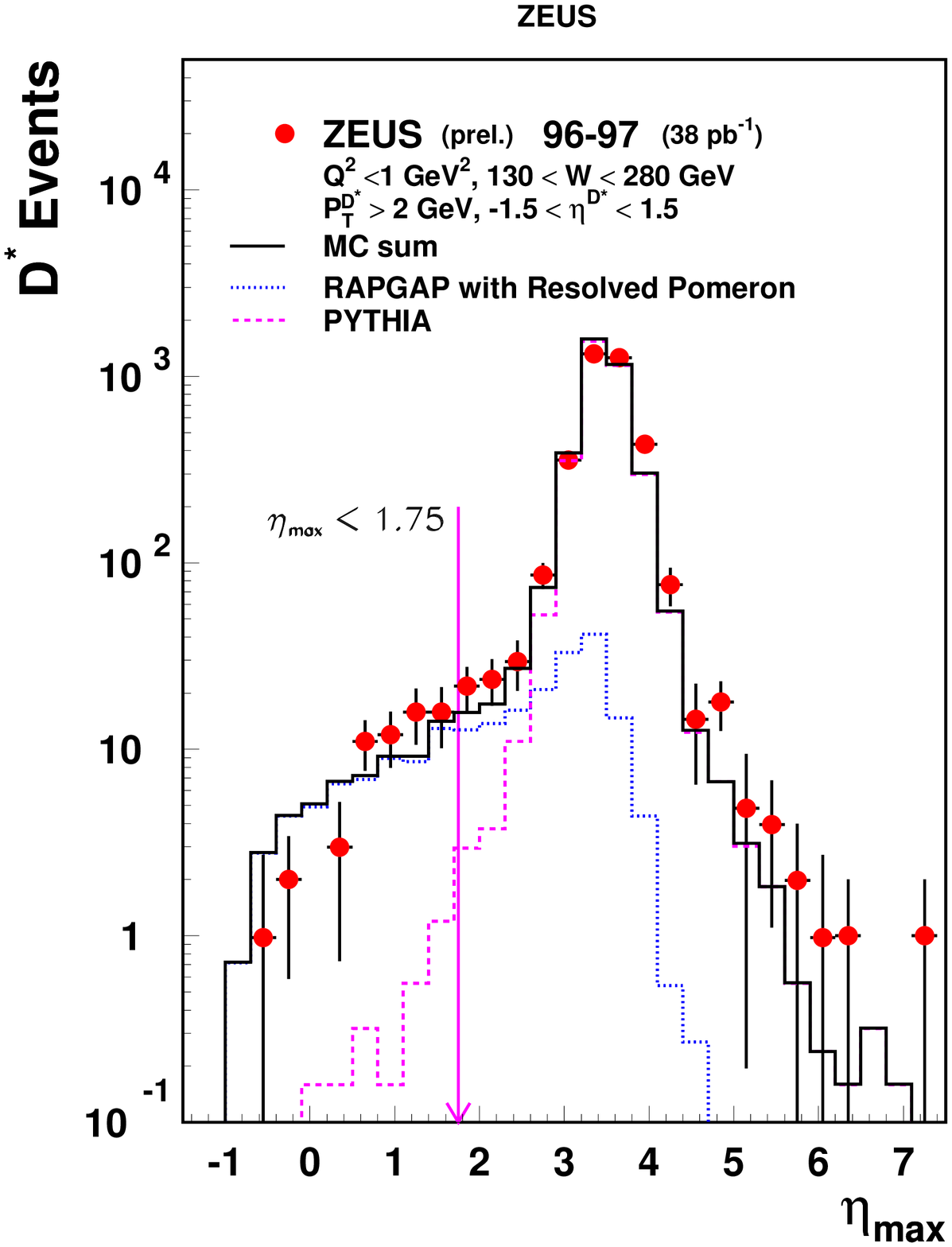,width=14cm}}}
\bf\caption{  
Comparison of the measured $\eta_{max}$ distribution (dots) with the sum of 
 the diffractive and non-diffractive MC distributions (histograms)  
for events with $D^*$ mesons. 
 $D^*$ candidates were selected in the kinematic region 
 $Q^2<1$ \GeVs, $130<W<280$ \Gev ,   
 $p_T^{D^*}> 2 $ \Gev ~and \mbox{$\vert\eta^{D^*}\vert< 1.5$}. 
Sum distribution of the diffractive resolved Pomeron RAPGAP MC (dotted histogram)
and non-diffractive MC  (dashed histogram) events were normalised  %
to have the same area as the data distribution.
  }
\label{fig:ds-etamxzf}
\end{figure}
%
%
\begin{figure}[htb]
\centerline{\hbox{
\psfig{figure=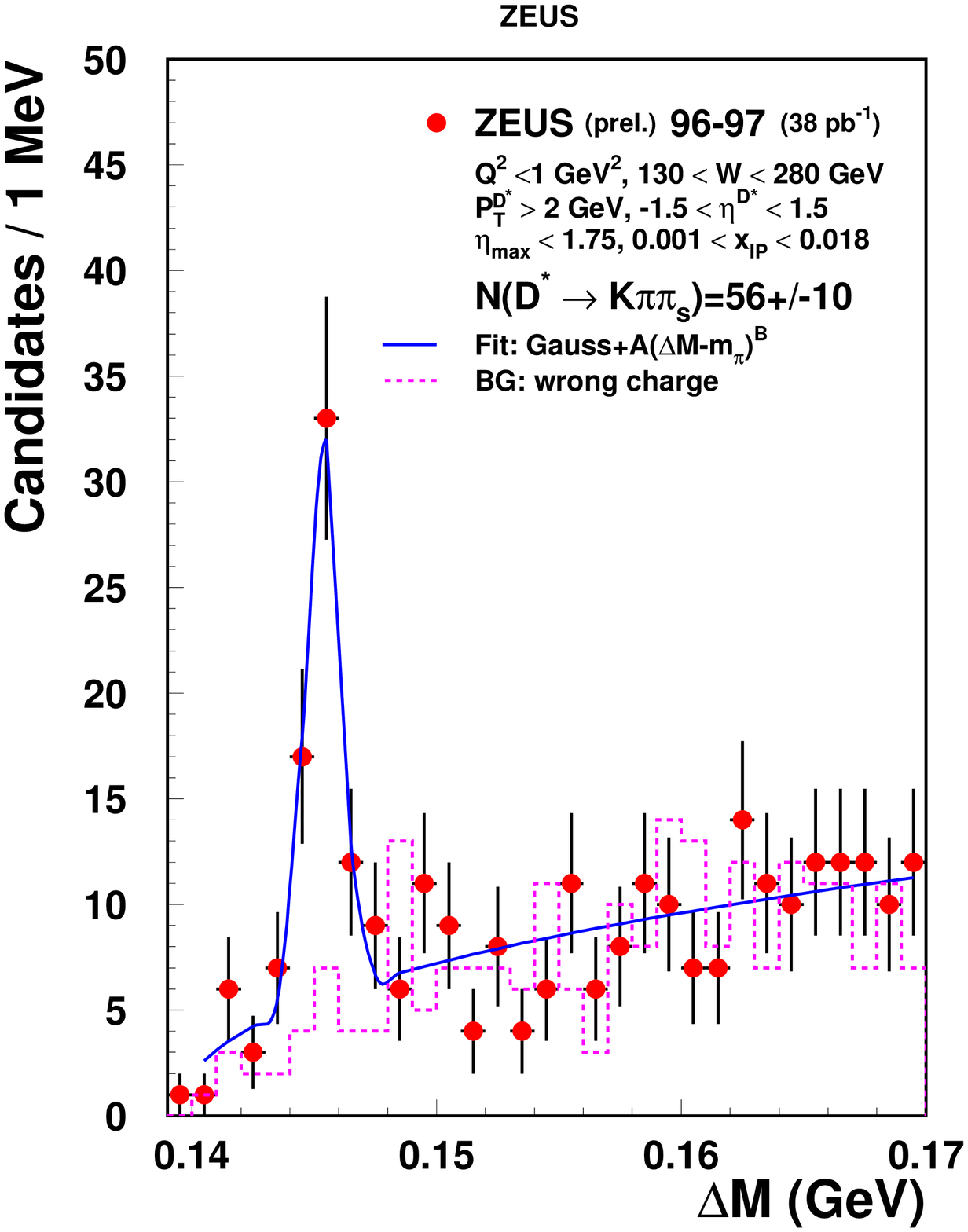,width=14cm}}}
\bf\caption{ 
The $\Delta M $distribution for the \(D^*\) diffractive photoproduction reaction with
$D^*\to(D^0\to K\pi)\pi_s$  for $Q^2<1$ \GeVs, $130<W<280$ ~\Gev and 
\mbox{$0.001<x_{\pomsub}<0.018$}. The kinematic range of measurements is
 $p_T^{D^*}> 2 $ \Gev ~and \mbox{$\vert\eta^{D^*}\vert< 1.5$}.
 The dots are for the right charge combinations, and the dashed  
histogram is for the wrong charge combinations from the $D^0$ signal region 
( 1.80 - 1.92 \Gev ). The full line is the result of a fit to a sum of a Gaussian
and the functional form $A (\Delta M-m_{\pi})^B$. %
}
\label{fig:ds-dm}
\end{figure}
\begin{figure}[tb]
\centerline{\hbox{
\psfig{figure=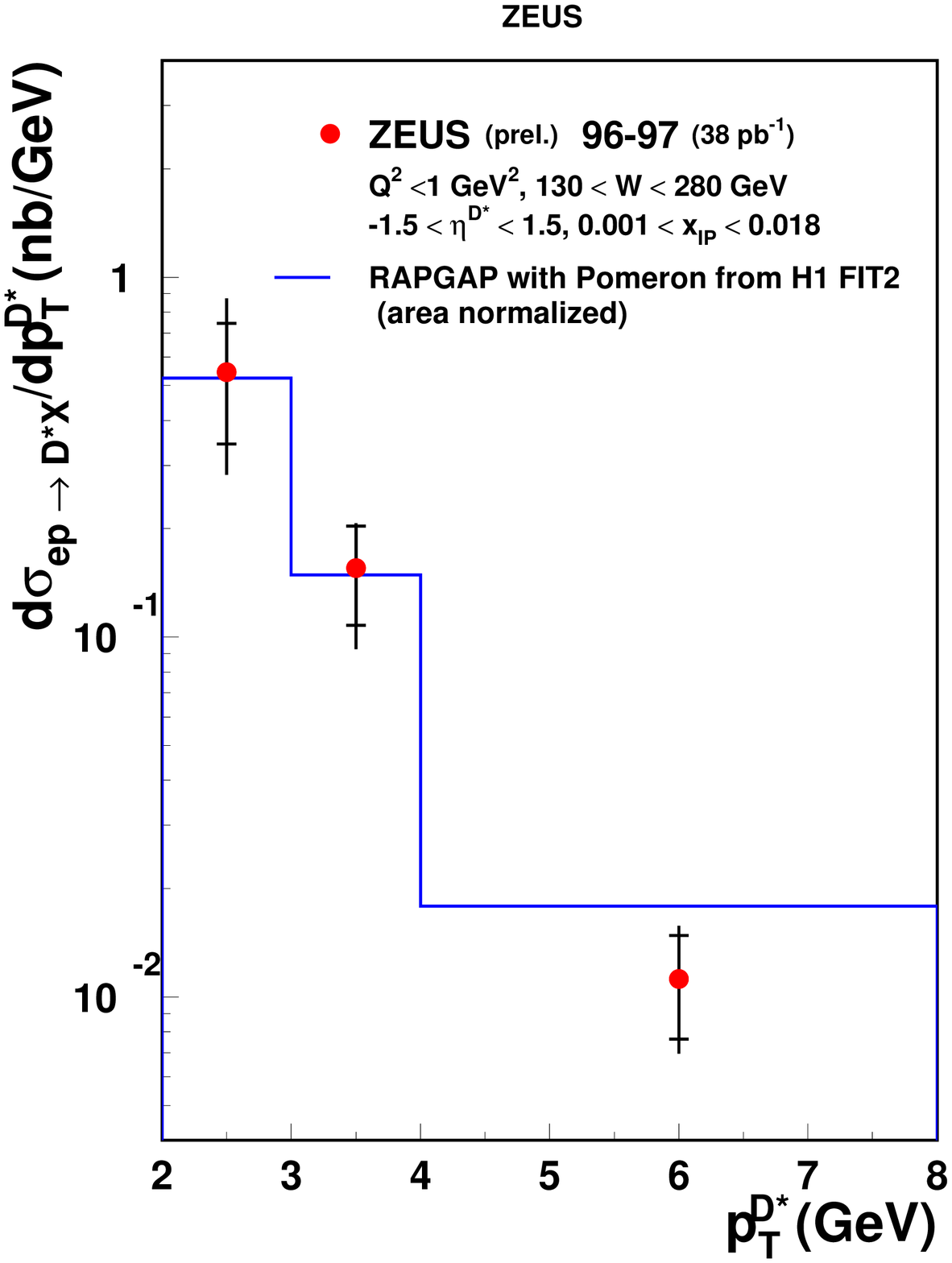,width=14cm}}} 
\bf\caption{   
Differential cross section ~${d\sigma} / {dp_T^{D^*}}$ (dots) 
for the diffractive photoproduction reaction 
\mbox{$ep\to e^{\prime}D^*Xp^{\prime}$} for \mbox{$Q^2<1$ \GeVs}, \mbox{$130<W<280$ \Gev} and 
\mbox{$0.001<x_{\pomsub}<0.18$}.
The kinematic range of measurements is 
\mbox{$p_T^{D^*}> 2$} \Gev ~and \mbox{$\vert\eta^{D^*}\vert< 1.5$}. 
The inner bars show the  statistical errors, and the outer bars correspond to 
 the statistical and systematic errors, added in quadrature. The data are compared 
with the distributions of the RAPGAP MC diffractive events, 
simulated in the framework of the resolved Pomeron model with the H1 FIT2 
Pomeron parametrization ( histogram).
The MC distribution has been normalized to have the same area
 as the data distribution.
}
\label{fig:ds-pt-xsec}
\end{figure}
\begin{figure}[tb]
\centerline{\hbox{
\psfig{figure=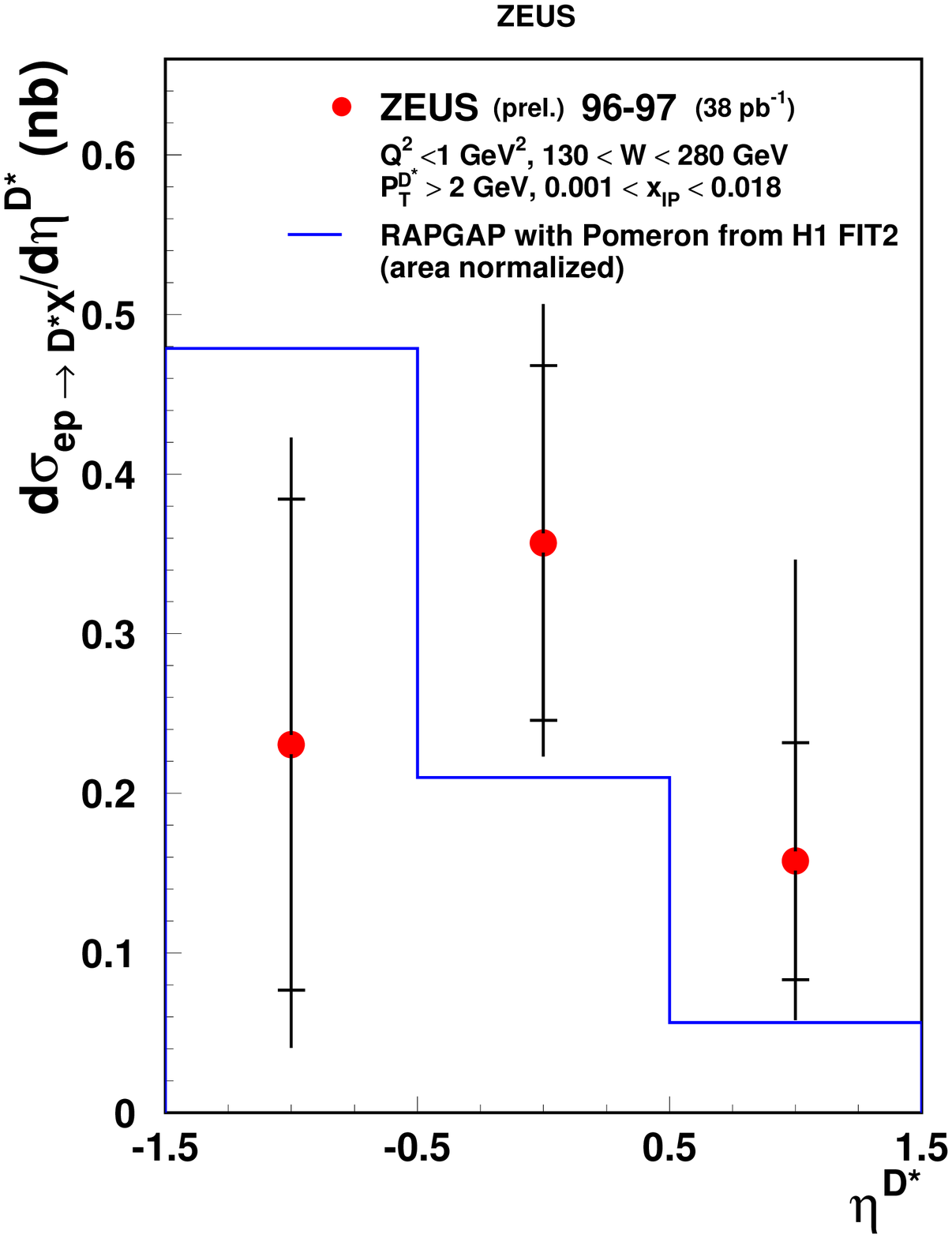,width=14cm}}} 
\bf\caption{  
Differential cross section ~${d\sigma} / {d\eta^{D^*}}$ (dots) 
for the diffractive photoproduction reaction 
\mbox{$ep\to e^{\prime}D^*Xp^{\prime}$} for \mbox{$Q^2<1$ \GeVs}, \mbox{$130<W<280$ \Gev} and 
\mbox{$0.001<x_{\pomsub}<0.18$}.
The kinematic range of measurements is 
\mbox{$p_T^{D^*}> 2 $} \Gev  ~and \mbox{$\vert\eta^{D^*}\vert< 1.5$}. 
The inner bars show  the statistical errors, and the outer bars correspond to 
 thestatistical and systematic errors, added in quadrature. The data are 
 compared with the distributions of the RAPGAP MC diffractive events, 
simulated in the framework of the resolved Pomeron model with the  H1 FIT2 
Pomeron parametrization ( histogram).
The MC distribution has been normalized to have the same area
 as the data distribution.
}
\label{fig:ds-eta-xsec}
\end{figure}
\begin{figure}[tb]
\centerline{\hbox{
\psfig{figure=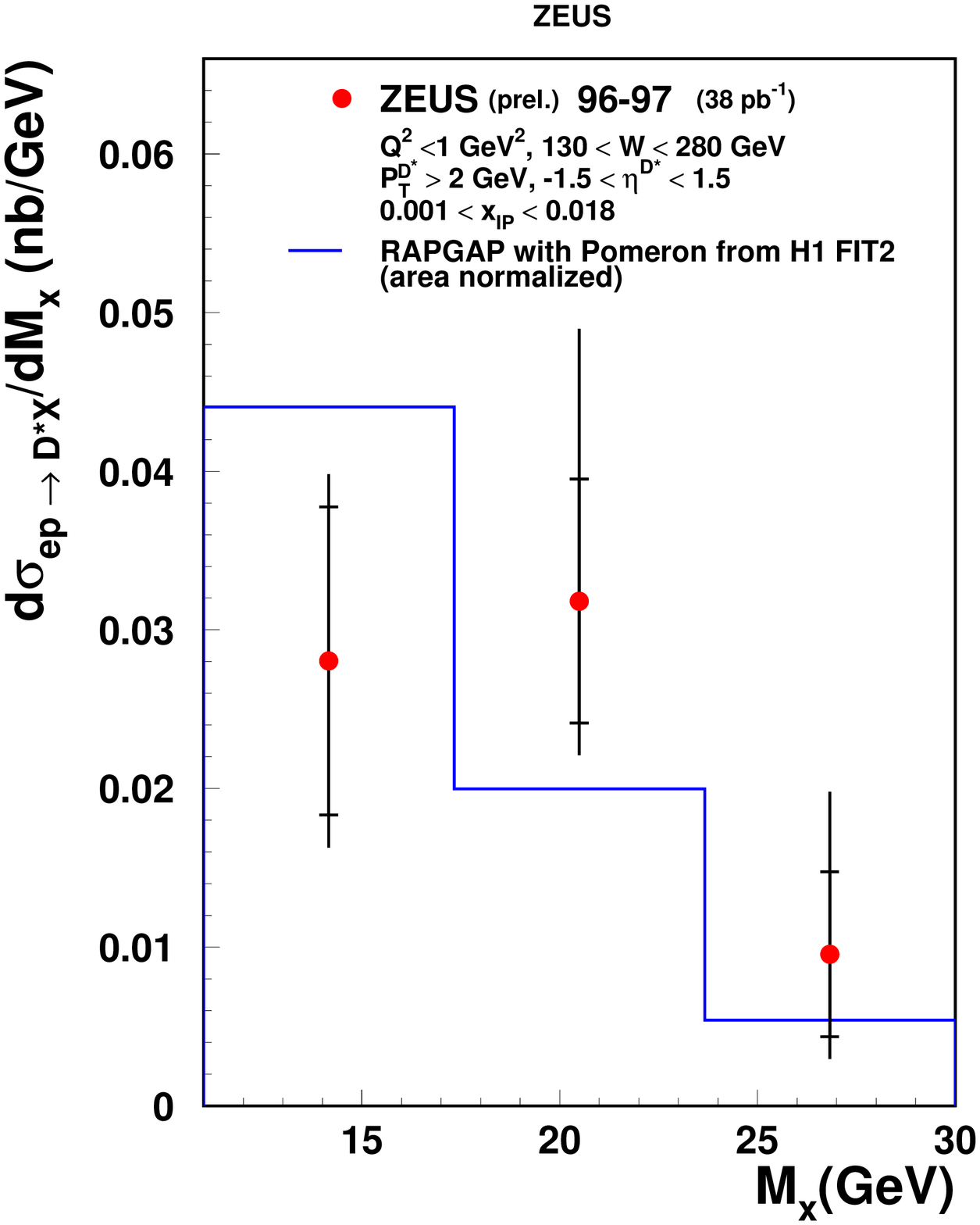,width=14cm}}} 
\bf\caption{   
Differential cross section ~${d\sigma} / {dM_X}$ (dots) 
for the diffractive photoproduction reaction 
\mbox{$ep\to e^{\prime}D^*Xp^{\prime}$} for \mbox{$Q^2<1$ \GeVs}, \mbox{$130<W<280$ \Gev} and 
\mbox{$0.001<x_{\pomsub}<0.18$}.
 The kinematic range of measurements is 
\mbox{$p_T^{D^*} > 2 $} \Gev ~and \mbox{$\vert\eta^{D^*}\vert< 1.5$}. 
The inner bars show the statistical errors, and the outer bars correspond to 
 thestatistical and systematic errors, added in quadrature. The data are 
 compared with the distributions of the RAPGAP MC diffractive events, 
simulated in the framework of the resolved Pomeron model with the H1 FIT2 
Pomeron parametrization ( histogram).
The MC distribution has been normalized to have the same area
 as the data distribution.
}
\label{fig:ds-mx-xsec}
\end{figure}
\begin{figure}[tb]
\centerline{\hbox{
\psfig{figure=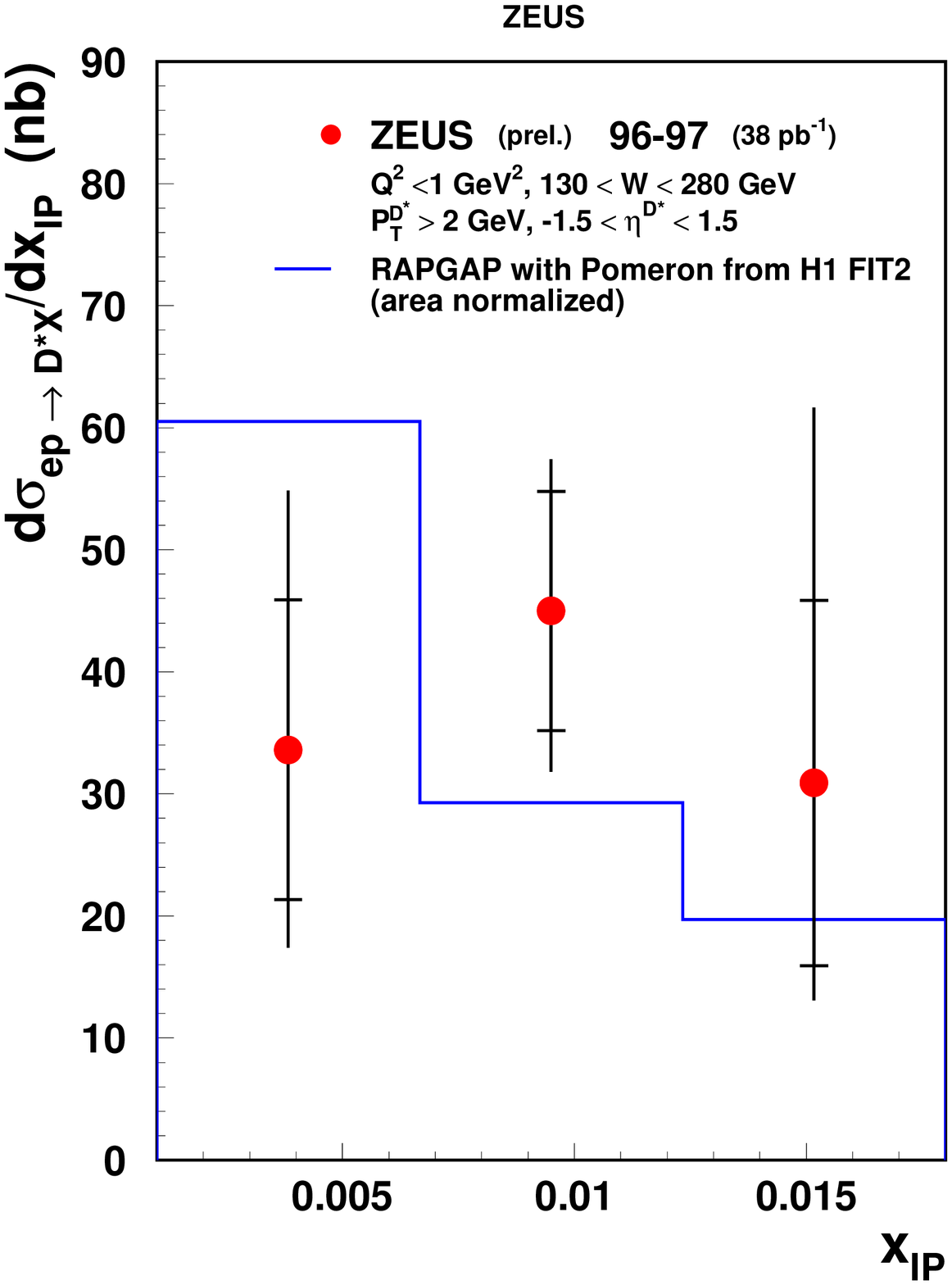,width=14cm}}} 
\bf\caption{  
Differential cross section ~${d\sigma} / {dx_{\pomsub}}$ (dots) 
for the diffractive photoproduction reaction 
\mbox{$ep\to e^{\prime}D^*Xp^{\prime}$} for \mbox{$Q^2<1$ \GeVs}, \mbox{$130<W<280$ \Gev} 
and \mbox{$0.001<x_{\pomsub}<0.18$}.
 The kinematic range of measurements is 
\mbox{$p_T^{D^*}> 2 $} \Gev ~and \mbox{$\vert\eta^{D^*}\vert< 1.5$}. 
The inner bars show the statistical errors, and the outer bars correspond to 
 the statistical and systematic errors, added in quadrature. The data are 
 compared with the distributions of the RAPGAP MC diffractive events, 
simulated in the framework of the resolved Pomeron model with  the H1 FIT2 
Pomeron parametrization ( histogram).
The MC distribution has been normalized to have the same area
 as the data distribution.
}
\label{fig:ds-xPom-xsec}
\end{figure}
\end{document}

%% file: com_def_tex/zeuspap_def.tex
\def\zero{{\scriptscriptstyle 0}}

\def\Z0{\ensuremath{Z^\zero}}

\def\Gev{\ensuremath{\text{Ge}@!@!@!@!\text{V\/}}}
\def\GeVs{\ensuremath{\text{Ge}@!@!@!@!\text{V\/}^2}}

\def\pbi{\,\text{pb}^{-1}}

\def\JB{{\scriptscriptstyle{\rm JB}}}

\def\SU2U1{{\rm SU}(2)\times{\rm U}(1)}

\mathchardef\qsm=63
\mathchardef\pls=43
\mathchardef\mns=512
\mathchardef\plm=518
\mathchardef\eql=61
\mathchardef\smallleft=300
\mathchardef\smallright=301
\mathchardef\perslsh=47
\mathchardef\les=316
\mathchardef\gre=318
\mathchardef\leq=532
\mathchardef\grq=533
\chardef\usc=95
\chardef\til=126

\def\sqr#1#2#3{{\vcenter{\hrule height.#3ex\hbox{\vrule width.#2ex height#1ex
    \kern#1ex\vrule width.#3ex}\hrule height.#2ex}}}

\def\angleto{\vrule width.035em height2.1ex depth-.56ex\unskip\kern-.6ex\to}
\def\perchc#1{{\raise.4ex\hbox{$\mkern4mu#1{\it\perslsh}_
             {\mkern-5mu\scriptscriptstyle{{\rm o}\!{\rm o}}}^
             {\mkern-12.8mu\scriptscriptstyle{\rm o}}$}}}

\catcode`\@=11 
\def\parenbar{\mathpalette\p@renb@r}
\def\p@renb@r#1#2{\vbox{%
  \ifx#1\scriptscriptstyle \dimen@.7em\dimen@ii.2em\else
  \ifx#1\scriptstyle \dimen@.8em\dimen@ii.25em\else
  \dimen@1em\dimen@ii.4em\fi\fi \offinterlineskip
  \ialign{\hfill##\hfill\cr
    \vbox{\hrule width\dimen@ii}\cr
    \noalign{\vskip-.3ex}%
    \hbox to\dimen@{$\mathchar300\hfil\mathchar301$}\cr
    \noalign{\vskip-.3ex}%
    $#1#2$\cr}}}
\catcode`\@=12 

\newbox\struttbox
\setbox\struttbox=\hbox{\vrule height1.65ex depth.485ex width0pt}
\def\strutt{\relax\ifmmode\copy\struttbox\else\unhcopy\struttbox\fi}
\def\stru#1#2{\relax\ifmmode\hbox{\vrule height#1 depth#2 width0pt}
\else\vrule height#1 depth#2 width0pt\fi}

\def\ronum#1{\uppercase\expandafter{\romannumeral#1}}
\def\ronuml#1{\expandafter{\romannumeral#1}}

\DeclareMathAlphabet{\mathbf}{OT1}{cmr}{bx}{sl}

%% file: com_def_tex/zeuspap_fmt.tex
\renewcommand{\baselinestretch}{1.12}

 {\end{list}}
 {\end{list}}
 {\end{list}}

\catcode`\@=11 
\newlength{\@fninsert}
\newlength{\@fnwidth}
\renewcommand{\@makefntext}[1]%
  {\noindent\makebox[\@fninsert][r]{\@makefnmark}\hfil%
  \parbox[t]{\@fnwidth}{#1}}
\catcode`\@=12 
\newlength{\localtextwidth}
\catcode`\@=11 
\newsavebox{\tmpbox}
\newlength{\@captionmargin}
\newlength{\@captionwidth}
\newlength{\@captionitemtextsep}
\renewcommand{\@makecaption}[2]%
  {\def\baselinestretch{0.95}%
   \vspace{10.pt}
   \setlength{\@captionwidth}{\localtextwidth}
   \addtolength{\@captionwidth}{-\@captionmargin}
   \sbox{\tmpbox}{{\bf #1:}{\it #2}}%
   \ifthenelse{\lengthtest{\wd\tmpbox > \@captionwidth}}%
   {\centerline{\parbox[t]{\@captionwidth}%
   {\tolerance=2000\normalsize%
    {\bf #1:}\hspace{\@captionitemtextsep}{\it #2}}}}%
   {\centerline{{\bf #1:}\kern1.em{\it #2}}}}
\catcode`\@=12 
\catcode`\@=11 
\renewcommand\section{\@startsection{section}{1}{\z@}%
                                   {-3.5ex \@plus -1ex \@minus -.2ex}%
                                   {2.3ex \@plus.2ex}%
                                   {\normalfont\Large\bfseries}}
\renewcommand\subsection{\@startsection{subsection}{2}{\z@}%
                                   {-3.25ex\@plus -1ex \@minus -.2ex}%
                                   {1.5ex \@plus .2ex}%
                                   {\normalfont\large\bfseries}}
\renewcommand\subsubsection{\@startsection{subsubsection}{3}{\z@}%
                                   {-3.25ex\@plus -1ex \@minus -.2ex}%
                                   {1.5ex \@plus .2ex}%
                                   {\normalfont\large\bfseries}}
\renewcommand\paragraph{\@startsection{paragraph}{4}{\z@}%
                                   {3.25ex \@plus1ex \@minus.2ex}%
                                   {1.2ex \@plus .2ex}%
                                   {\normalfont\normalsize\bfseries}}
\catcode`\@=12 

\newsavebox{\sesbox}
\newlength{\seslen}

%% file: com_def_tex/charm_com.tex
\newcommand{\Ds}         {\mbox{$D^{\ast}$}}

\def\et10{ E_{\perp}^{\theta > 10^\circ}}

%% file: RASproc00hepex.bbl
\begin{thebibliography}{99}
%
  \bibitem{ZEUS-D*-incl}
%
ZEUS Collaboration (M.Derrick {\it et al}.), Phys. Lett. B {\bf 349}, 225(1995);
%
ZEUS Collaboration (J.Breitweg {\it et al}.), Phys. Lett. B {\bf 401}, 192(1997);
                                      Phys. Lett. B {\bf 407}, 402(1997);
                                      Eur. Phys. J. C {\bf 6}, 67(1999);
                                      Eur. Phys. J. C  {\bf 12}(1), 35(2000).
%
  \bibitem{H1-D*-incl} 
%
H1 Collaboration (S.Aid {\it et al}.),    Nucl. Phys. B {\bf 472}, 32(1996);
                                           Z. Phys. C {\bf 72}, 593(1996);
%
H1 Collaboration (C. Adlof {\it et al}.),   Nucl. Phys. B {\bf 545}, 21(1999). 

%
  \bibitem{ZEUS-D*-disdif}
 ZEUS Collaboration (paper N-645), {\it Int. Europhys. Conf. on High Energy 
 Physics,  Jerusalem, Israel,1997};
ZEUS Collaboration, (paper N-527), {\it Int. Europhys. Conf. on High Energy 
Physics,  Tampere, Finland,1999};
 ZEUS Collaboration {\it 8th International Workshop on Deep-Inelastic 
 Scattering DIS2000, Liverpool,  England, 2000};
 ZEUS Collaboration (abstract 874), {\it the XXXth Int. Conf. on  
                      High Energy Physics, Osaka, Japan, 2000.}
%
  \bibitem{H1-D*-disdif}
  H1 Collaboration (pa02-60), {\it 28th Int. Conf. on High Energy Physics, 
                                               Warsaw, Poland, 1996};
  H1 Collaboration, {\it 29th Int. Conf. on High Energy Physics ICHEP98,
                                               Vancouver,  Canada, 1998};
  H1 Collaboration, {\it 8th International Workshop on Deep-Inelastic 
  Scattering DIS2000, Liverpool, 
  England, 2000.}
  
%
  \bibitem{ZEUS} 
ZEUS Collaboration (M.Derrick {\it et al}.), Phys. Lett. B {\bf 293}, 
465(1992); \\ The ZEUS detector: Status Report 1993, DESY, 1993.
%
  \bibitem{CTD} 
 N.Harnew {\it et al}., Nucl. Instrum. Methods A {\bf 279}, 290(1989); \\
 B.Foster {\it et al}., Nucl. Phys. Proc. Suppl. B {\bf 32}, 181(1993),
  ~Nucl. Instrum. Methods A {\bf 338}, 254(1994).
%
  \bibitem{CAL} 
M.Derrick {\it et al}., Nucl. Instrum. Methods A, {\bf 309}, 77(1991);
A.Andresen {\it et al}., ibid., 101(1991);
A.Caldwell {\it et al}.,Nucl. Instrum. Methods,  A {\bf 321}, 356(1992);
A.Bernstein {\it et al}., Nucl. Instrum. Methods, A {\bf 336}, 23(1993). 
%
  \bibitem{LUMI} 
J.Andruszk\'{o}w {\it et al}., DESY 92-066, 1992; \\
ZEUS Collaboration (M.Derrick {\it et al})., Z. Phys. C {\bf 63}, 391(1994);
                                 Nucl. Phys. B {\bf 303}, 634(1988).  
%
  \bibitem{EFO} 
ZEUS Collaboration (M.Derrick {\it et al}.), Eur. Phys. J. C {\bf 1}, 81(1998);
Eur. Phys. J. C {\bf 6}, 43(1999); G.Briskin, PhD Thesis, University of 
Tel Aviv, 1998. 
%
  \bibitem{JB} 
F.Jacquet and A.Blondel, {\it Proc. of the Study for an $ep$ Facility 
for Europe}, DESY 79-48, 391(1979).
%
  \bibitem{no-electron} 
  ZEUS Collaboration (M.~Derrick {\it et al}.), Phys. Lett. B {\bf 322},  287(1994).
%
  \bibitem{PDG} 
  C.~Caso {\it et al}., Particle Data Group, Eur. Phys. J. C {\bf 3}, 1(1998).
%
  \bibitem{RAPGAP}
  H.Jung, Comp. Phys. Comm.  {\bf 86}, 147(1995). 
%
  \bibitem{PYTHIA}  
  T.~Sj\"{o}strand, Comp. Phys. Comm. {\bf 82},  74(1994).
%
  \bibitem{HERWIG} 
  G.Marchesini {\it et al}., Comp. Phys. Comm. {\bf 67}, 465(1992). 
%
  \bibitem{IPres} 
 G.Ingelman and P.Schlein, Phys. Lett. B {\bf 152}, 256(1985); \\
  A.Donnachie and P.V.Landshoff, Nucl. Phys. B {\bf 303}, 634(1988). 
%
  \bibitem{IPfitH1}
  H1 Collaboration (C.Adloff {\it et al}.), Z. Phys. C {\bf 76}, 613(1997). 
%
  \bibitem{MRSG}
A.D.~Martin, W.J.~Stirling, and R.G.~Roberts, Phys. Lett. B {\bf 354}, 155(1995).
%
  \bibitem{GRV}
  M.~Gl\"{u}ck, E.~Reya, and A.~Vogt, Phys. Rev. D {\bf 46}, 1973(1992).
%
  \bibitem{LUND} 
  T.~Sj\"{o}strand, Comp. Phys. Comm. {\bf 39}  347(1986); \\
  T.~Sj\"{o}strand and M.Bengtsson, Comp. Phys. Comm. {\bf 43},  367(1987).
%
  \bibitem{prodiss}
ZEUS Collaboration (J.Breitweg {\it et al}.), Phys. Lett. B {\bf 315}, 43(1999);
                                             Eur. Phys. J. C {\bf 1}, 81(1998).
  \bibitem{c-suppr} 
                N.N.Nikolaev and B.G.Zakharov, Z. Phys. C {\bf 53}, 331(1992).
%
  \bibitem{HJung} H.Jung, private communication.
%
%
\end{thebibliography}
